\documentclass[aps,prl,twocolumn,showpacs,superscriptaddress]{revtex4-1}
\usepackage[active]{srcltx}
\usepackage{amsthm,mathrsfs}
\usepackage{amssymb,amsmath}
\usepackage{xargs}
\usepackage{xifthen}
\usepackage{graphicx}  

\def\ie{{i.~e.}}
\mathchardef\minus="002D

\def\<{\langle}\def\>{\rangle}
 \def\ket#1{| #1 \rangle} 
 
\def\ketbra#1#2{| #1 \rangle \langle#2 |}

\def\Z{\mathbb Z} 
 
\def\Reals{\mathbb{R}}

\def\L2{{\mathcal L}_2}

\def\M{\mathcal D}
\def\B{{L}^D_\beta}
\def\L{L_\beta}
\def\d{\operatorname{d}\!}

\def\df#1#2 {\!\!\frac{\mathop{\mathrm{d}#1}}{#2}\,}
\def\ddf#1#2 {\!\!\frac{\mathop{\mathrm{d}#1}}{\mathrm{d}#2}\,}

\begin{document}

\title{Doubly-Special Relativity from Quantum Cellular Automata}

\author{Alexandre Bibeau-Delisle}\email[]{bibeauda@iro.umontreal.ca}
\affiliation{Universit\'e de Montr\'eal, DIRO}
 \author{Alessandro Bisio} \email[]{alessandro.bisio@unipv.it}
\affiliation{Dipartimento di Fisica dell'Universit\`a di Pavia, via
  Bassi 6, 27100 Pavia} \affiliation{Istituto Nazionale di Fisica
  Nucleare, Gruppo IV, via Bassi 6, 27100 Pavia}
\author{Giacomo~Mauro~D'Ariano} \email[]{dariano@unipv.it}
\affiliation{Dipartimento di Fisica dell'Universit\`a di Pavia, via Bassi 6, 27100 Pavia} \affiliation{Istituto Nazionale di Fisica Nucleare, Gruppo IV, via Bassi 6, 27100 Pavia} 
\author{Paolo Perinotti}\email[]{paolo.perinoti@unipv.it}
\affiliation{Dipartimento di Fisica dell'Universit\`a di Pavia, via Bassi 6, 27100 Pavia} \affiliation{Istituto Nazionale di Fisica Nucleare, Gruppo IV, via Bassi 6, 27100 Pavia} 
\author{Alessandro Tosini}\email[]{alessandro.tosini@unipv.it}
\affiliation{Dipartimento di Fisica dell'Universit\`a di Pavia, via
  Bassi 6, 27100 Pavia} \affiliation{Istituto Nazionale di Fisica
  Nucleare, Gruppo IV, via Bassi 6, 27100 Pavia}

\begin{abstract}
  It is shown how a Doubly-Special Relativity model can emerge from a quantum cellular automaton
  description of the evolution of countably many interacting quantum systems. We consider a
  one-dimensional automaton that spawns the Dirac evolution in the relativistic limit of small
  wave-vectors and masses (in Planck units). The assumption of invariance of dispersion relations
  for boosted observers leads to a non-linear representation of the Lorentz group on the
  $(\omega,k)$ space, with an additional invariant given by the wave-vector $k=\pi /2$. The
  space-time reconstructed from the $(\omega,k)$ space is intrinsically quantum, and exhibits the
  phenomenon of {\em relative locality}.
\end{abstract}

\maketitle

The existence of a fundamental scale of length or mass, which can be identified with the Planck
scale, is a ubiquitous feature of quantum gravity models
\cite{polchinski1996tasi,polchinski2003string,rovelli1990loop,ashtekar1992weaving,rovelli1995discreteness,carlip2001quantum,garay1995quantum}.
The appearance of the minimum length $ \ell_P=\sqrt{\hbar G/c^3}$ is the result of combining the
fundamental constants that characterize physical theories describing different scales: $\hbar$
(quantum mechanics), $c$ (special relativity), and $G$ (gravity). The so-called {\em Planck length}
$\ell_P$ is commonly regarded as the threshold below which the intuitive description of space-time
breaks down, and new phenomenology is expected. A natural hypothesis is that quantum features become
crucial in determining the structure of space-time below the Planck scale, leading to a radical
departure from the traditional geometric concepts.  This perspective makes one wonder about the
fate of Lorentz symmetry at the Planck scale. A possible way of tackling this question is to
consider a theory with two observer-independent scales, the speed of light and the Planck length, as
proposed in the models of Doubly-Special Relativity (DSR)
\cite{amelino2002relativity,amelino2001planck,amelino2002quantum,amelino2001testable,PhysRevLett.88.190403,magueijo2003generalized}.
All the DSR models share the feature of a non-linear deformation of the Poincar\'e symmetry that
eventually leads to a modification of the quadratic invariant
\begin{align}
  \label{eq:usual-disp}
  E^2 = p^2 + m^2.
\end{align}
Such deformed kinematics are especially interesting since they provide new phenomenological
predictions, e.~g.~wavelength dependence of the speed of light and a modified threshold for particle
creation in collision processes.  Evidences for a violation of the Lorentz energy-momentum
dispersion relation \eqref{eq:usual-disp} have recently been sought in astrophysics, see
e.~g.~the thresholds for ultra-high-energy cosmic rays
\cite{takeda1998extension,finkbeiner2000detection}, and in cold-atom experiments
\cite{PhysRevLett.103.171302}.

A recent approach to a Planck scale description of physical kinematics is that of quantum cellular
automata (QCAs) \cite{veryoriginal,darianopla,BDTqcaI,d2013derivation,bisio2013dirac}.  The QCA
generalizes the notion of cellular automaton of von Neumann \cite{neumann1966theory} to the quantum
case, with cells of quantum systems interacting with a finite numer of neighbors via a unitary
operator describing the single step evolution \footnote{ The general theory of QCA is rigorously
  treated in Refs.  \cite{arrighi2011unitarity,gross2012index} while first attempts to model
  relativistic dynamics with QCA appears in \cite{bialynicki1994weyl,meyer1996quantum}.}.  We assume
that each cell $x$ of the lattice corresponds to the local value $\psi(x)$ of a quantum field whose
dynamics is described by a QCA.  From this perspective the usual quantum field evolution should
emerge as a large scale approximation of the automaton dynamics occurring at an hypothetical
discrete Planck scale. In Ref.~\cite{veryoriginal} a QCA--called {\em Dirac QCA} in the
following--has been proposed for describing the Planck-scale physics of the Dirac field in $d=1$
space dimension, assuming the Planck length as the distance between the cells. In Ref.
\cite{BDTqcaI} it has been shown that the dynamics of such QCA recovers the usual Dirac evolution in
the relativistic limit of small wave-vectors $ k \ll 1 $ and small masses $m \ll 1 $ (everything
expressed in Planck units). In Ref.~\cite{d2013derivation} it has been shown that in $d=3$ space
dimensions and for minimal number of field components only two QCAs satisfy locality, homogeneity,
and isotropy of the of the quantum-computational network, the two QCAs being connected by CPT
symmetry, and giving the Dirac evolution in the relativistic limit. In $d=1,2$ space dimensions
there is instead only one QCA satisfying the above requirements.

Clearly the QCA theoretical framework cannot enjoy a continuous Lorentzian space-time, along with
the usual Lorentz covariance, which must break down at the Planck scale. For this reason the notion
itself of a boosted reference frame as based on an Einsteinian protocol has still to be refined.
However, whatever the final physical interpretation of the relativity principle, it must
include the invariance of the dispersion relation in any of its expressions, being at the core of the physical law. In this Letter we explore this route and, assuming
the invariance of the Dirac QCA dispersion relation, we find a non-linear representation of the
Lorentz group which exhibits the typical features of a DSR with an invariant energy scale. For the
present purpose without loss of generality we focus on the easiest $d=1$ dimensional case. As for
any DSR model, it turns out that the space-time emerging from the automaton exhibits {\em relative
  locality}, namely the phenomenon according to which the coincidence of two events is
observer-dependent \cite{schutzhold2003large,amelino2011relative,PhysRevD.84.084010}. Specifically,
we will show that in the automaton case the coincidence of particles trajectories is no longer
observer independent. Contrarily to the usual special relativity, where the Poincar\'e group acts
linearly both in the position and in the momentum space, in the DSR scenario there are essentially
no restrictions on the non-linear energy-momentum transformations, allowing for a variety of
possible models. Since these models are generally inequivalent from the physical point of view, an
open problem is to single out one of them via physical principles.  The quantum cellular automaton
provides a microscopic dynamical model which naturally introduces a DSR.

The Dirac QCA describes the one step evolution $\psi(x)\rightarrow \boldsymbol{\mathrm U}\psi(x)$ of
a two components field $\psi(x) :=(\psi_r(x),\psi_l(x))^T$ defined on a discrete array $x\in\Z$ of
quantum cells, with $\psi_r$ and $\psi_l$ denoting the left and right field modes. As proved in
Ref.~\cite{meyer1996quantum} for $d=1$ and in Ref. \cite{d2013derivation} for any $d$, in a non
interacting scenario all scalar fields exhibit trivial evolution and the minimal internal dimension
for the free field is two. The one dimensional Dirac QCA can be derived by imposing the invariance
with respect to the symmetries of the causal network \cite{BDTqcaI,d2013derivation}, and is given by
\begin{equation}\label{eq:evol}
\boldsymbol{\mathrm U}=\begin{pmatrix}
  n S &-im\\
  -im & n S^\dag
\end{pmatrix},\quad n^2+m^2=1,
\end{equation}
with $S$ the shift operator $S\psi(x) :=\psi(x+1)$. The canonical basis of the Fock space of the field
states are obtained by applying to the vacuum state $\ket{\Omega}$ the creation operators
$\psi^\dag_{s}(x)$, $s=r,l$. In the following we restrict ourselves to the one-particle sector for which an
orthonormal basis is given by the states $\ket{s}\ket{x}:=\psi^\dag_s(x)\ket{\Omega}$. We write
a generic one-particle state as $\ket{\psi}=\sum_{x,s}g_s(x)\ket{s}\ket{x}$ and Eq.~\eqref{eq:evol}
defines a unitary matrix $U$ on $\mathbb{C}^2 \otimes l_2(\mathbb Z)$. In the Fourier transformed
basis $\ket{s}\ket{\phi(k)}$, with $\ket{\phi(k)}:=(2\pi)^{-1/2}\sum_{x} e^{-ikx}\ket{x}$, $k\in
B:=[-\pi,\pi]$, the matrix $U$ is written as
\begin{equation}
U=
\int_B\!\! \mathrm{d}k\, \hat U(k)\otimes\ketbra{\phi(k)}{\phi(k)},\quad \hat U(k) =
\begin{pmatrix}
    n e^{ik} & -i m\\
-i m & n e^{-ik}
\end{pmatrix}, \nonumber
\end{equation}
whose eigenvalues are $\exp[\pm i \omega(k)]$ where 
the function $\omega(k)$ is given by
\begin{align}\label{eq:disp}
\cos^2 \omega =(1-m^2)\cos^2 k,
\end{align}
which is the dispersion relation of the Dirac automaton. In the limit of small wave-vectors and
masses Eq.~\eqref{eq:disp} reduces to $\omega^2=k^2+m^2$, and we recover the Lorentz dispersion
relation of Eq. \eqref{eq:usual-disp} \footnote{For large wave-vectors a thorough interpretation of
  $k$ and $\omega$ in terms of the momentum and energy, respectively, would need a full interacting
  theory.}.  Disregarding the internal degrees of freedom, we consider the dispersion relation in
Eq.  \eqref{eq:disp} as the core dynamics of the theory which should be independent of the reference
frame. In one spatial dimension the Lorentz group consists in only the boost transformations which
in the energy-momentum sector are represented by the linear map
\begin{equation}\label{eq:Lorentz-boosts}
  \L:
  (\omega,k)\mapsto(\omega',k')=\gamma(\omega-\beta
  k,k-\beta\omega),
\end{equation}
with $\gamma:=(1-\beta^2)^{-1/2}$. It is immediate to
check that the automaton dispersion relation of Eq. \eqref{eq:disp} is not
invariant under such standard boosts.

Following the DSR proposal of preserving the Lorentz group structure,
the linear Lorentz boosts in Eq. \eqref{eq:Lorentz-boosts} should be replaced
by a non-linear representation of the kind
\begin{equation}\label{eq:non-linear-boosts}
  \B:= \M^{-1}\circ\L\circ\M,
\end{equation}
where $\M:\Reals^2\to\Reals^2$ is a non-linear map.

The specific form of $\M$ gives rise to a particular energy-momentum Lorentz deformation. As pointed
out in \cite{magueijo2003generalized}, in order to realize a DSR model, the non linear map $\M$, has
to satisfy the following constraints: {\em i}) the Jacobian matrix $J_{\M}(k,\omega)$ of $\M$
evaluated in $k=\omega=0$ must be the identity, ensuring that the non-linear transformations $\B$
recover the standard boosts in the regime of small momenta and energies; {\em ii}) since $\L$ ranges
over the whole set $[0,\infty]\times[-\infty,+\infty]$, this set must be included in the
invertibility range of $\M$; {\em iii}) the model will exhibit an invariant energy scale only if the
map $\M$ has a singular point, namely some energy $\omega_{inv}$ which is mapped to $\infty$.
Restating Eq.~\eqref{eq:disp} in the following way 
\begin{align}
\frac{\sin^2\omega}{\cos^2 k}-\tan^2 k=m^2,
\nonumber
\end{align}
the non linear map $\M$ in \eqref{eq:non-linear-boosts} can be taken to be
\begin{align}\label{eq:map}
  \M:(\omega,k)\mapsto \M(\omega,k):=\left(\sin\omega/\cos k,\tan k\right).
\end{align}
One can show that the map in Eq. \eqref{eq:map} automatically satisfies the aforementioned
requirements {\em i})-{\em iii}) with the invariant energy $\omega_{inv}=\pi/2$.
By inserting the map \eqref{eq:map} into
Eq.~\eqref{eq:non-linear-boosts} we obtain the following deformed
Lorentz transformations
\begin{equation}\label{eq:extrans}
  \begin{split}
    \omega'&=\arcsin\left[\gamma\left(\sin\omega/\cos k-\beta\tan
        k\right)\cos k'\right],\\
    k'&=\arctan\left[\gamma\left(\tan k -\beta \sin\omega/\cos k\right)\right].
  \end{split}
\end{equation}
which leave the automaton dispersion relation of Eq. \eqref{eq:disp} invariant.

\begin{figure}[t]
\includegraphics[width=0.235\textwidth]{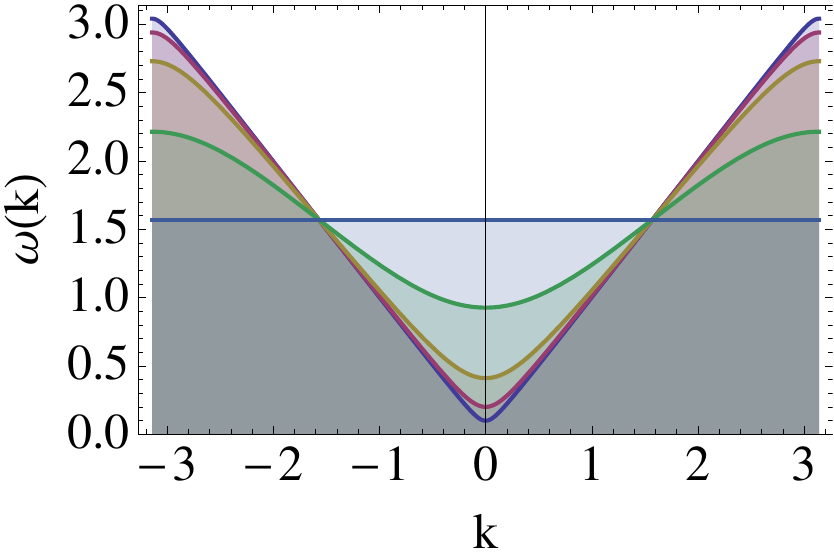}
\includegraphics[width=0.235\textwidth]{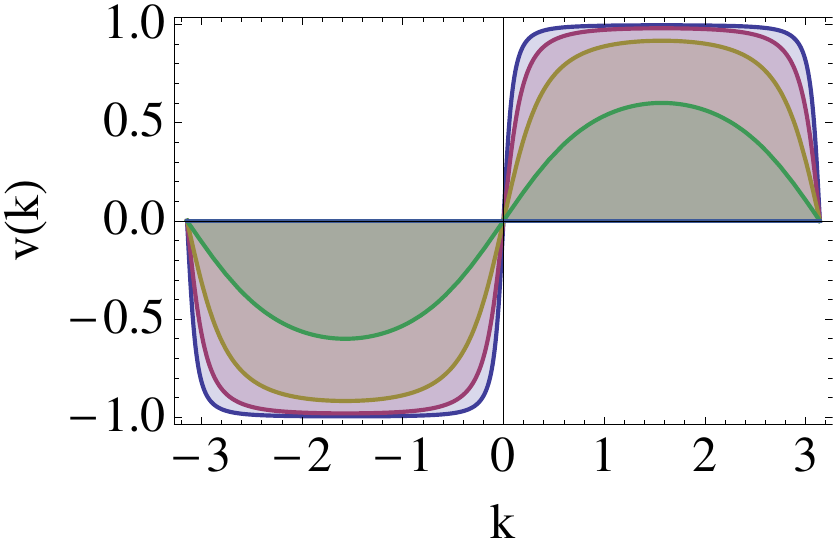}
\caption{(Colors online) The automaton dispersion relation (left) and group velocity (right) for $m=0.1,\,0.2,\,
  0.4,\,0.8,\,1$, from bottom to top at $k=0$ (left), and at $k=\pi/2$ (right).}
\label{fig:disp-group}
\end{figure}

The modified transformations have two symmetrical invariant
momenta $k=\pm\pi/2$ corresponding to the invariant energy $\omega_{inv}=\pi/2$
independently of $m$. The fixed points split the domain $B=[-\pi,\pi]$
into two regions $B=B_1\cup B_2$, with $B_1:=[-\pi/2,\pi/2]$ and
$B_2:=[-\pi,-\pi/2]\cup[\pi/2,\pi]$, which remain separate under all
possible boosts.  The points $k=\pm\pi/2$ correspond to maxima of
the group velocity $v:=\partial_k \omega(k)$ (see
Fig.~\ref{fig:disp-group}). While in region $B_1$ an increasing $k$
corresponds to an increasing group velocity, in region $B_2$ we see
the opposite behavior. However, as one can verify using the
transformations \eqref{eq:extrans}, a boosted observer who sees an
increased group velocity in $B_1$ also sees an increased group
velocity in $B_2$ since in both cases the momentum $k$ is mapped
closer to the invariant point. Since the two physical regions $B_1$
and $B_2$ exhibit the same kinematics they are indistinguishable in a
non interacting framework. For massless particles the Dirac automaton
dispersion relation \eqref{eq:disp} coincides with the undistorted one
$\omega^2=k^2$ and the group velocity no longer depends on $k$. Thus
the model we are considering does not exhibit a momentum-dependent
speed of light.

\begin{figure}[h!]
\includegraphics[width=0.9\columnwidth]{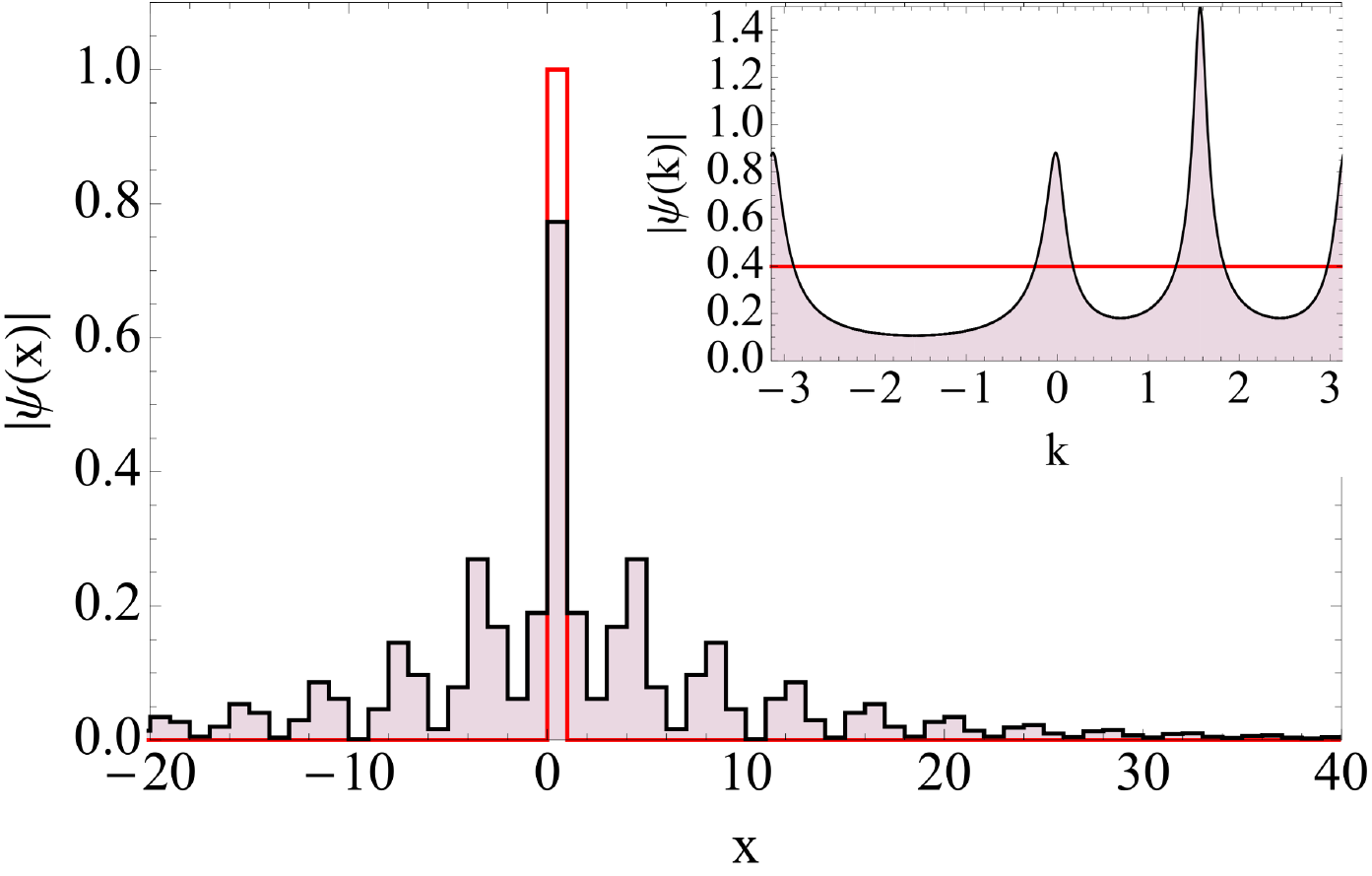}\\
\includegraphics[width=0.45\columnwidth]{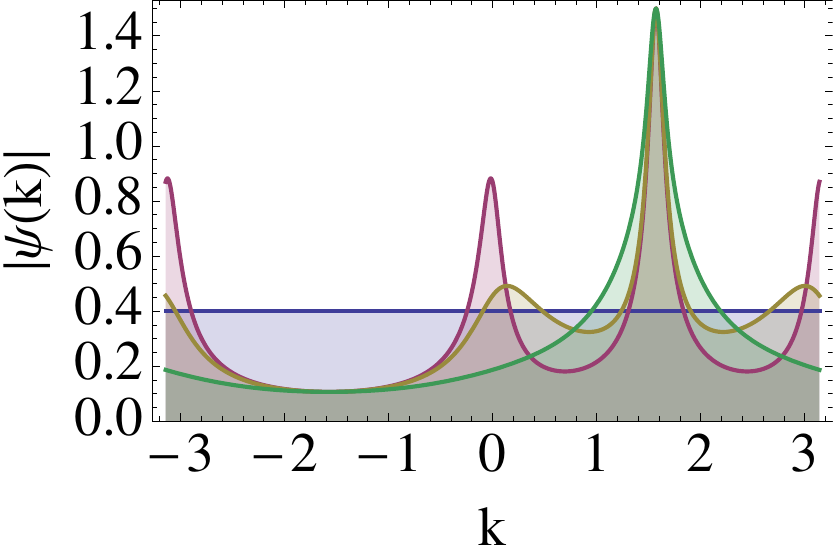}
\includegraphics[width=0.45\columnwidth]{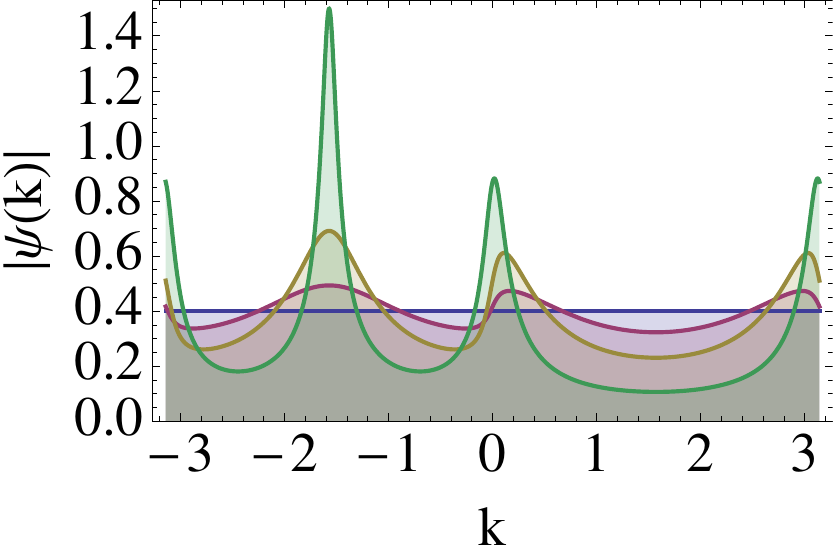}
\caption{(Colors online) {\bf
    Top figure:} Delocalization of a state localized at $x=0$ after a
  boost with $\beta=-0.99$ for mass $m=0.1$. {\bf Bottom figure:}
  Left: momentum representation of a boosted localized state for
  different values of the mass $m=0.1$ (red) $0.3$ (orange) $0.8$
  (green) with $\beta=-0.99$. Right: momentum representation of a
  boosted localized state for different values of the boost
  $\beta=0.4$ (red) $0.8$ (orange) $0.99$ (green) with
  $m=0.1$.}\label{fig:localized-state-boost}
\end{figure}

The action of the boosts \eqref{eq:extrans} on the states of the
automaton (disregarding the internal degrees of freedom)
reads
\begin{align}\label{eq:boost-automaton}
  \begin{split}
 \ket{\psi}=\int \! \! \d k{\mu}(k)\hat{g}(k)
 \ket{k}\xrightarrow{\B} &\int\!\d k{\mu}(k)\, \hat{g}(k)
 \ket{k'}   =\\
& =
\int \! \! \d k{\mu}(k')\, \hat{g}(k(k'))
 \ket{k'}   
  \end{split}
\end{align}
where $\mu(k)=[2(1-m^2)\tan\omega(k)]^{-1} $ is the density of the
invariant measure in the $k$-space, $k'$ is as in
Eq.~\eqref{eq:extrans}, and $\ket{k}:=(2(1-m^2) \tan \omega(k))^{1/2}
\ket{\phi(k)}$. One can verify that the transformation
\eqref{eq:boost-automaton} is unitary.  In
Fig.~\ref{fig:localized-state-boost} we show how a perfectly localized
state transforms under boosts.

\begin{figure}[b]
\includegraphics[width=0.23\textwidth]{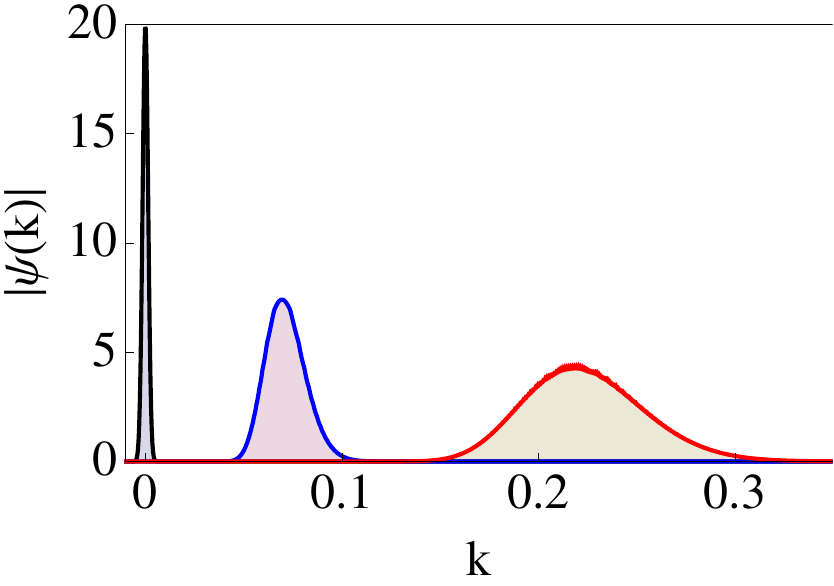}
\includegraphics[width=.24\textwidth]{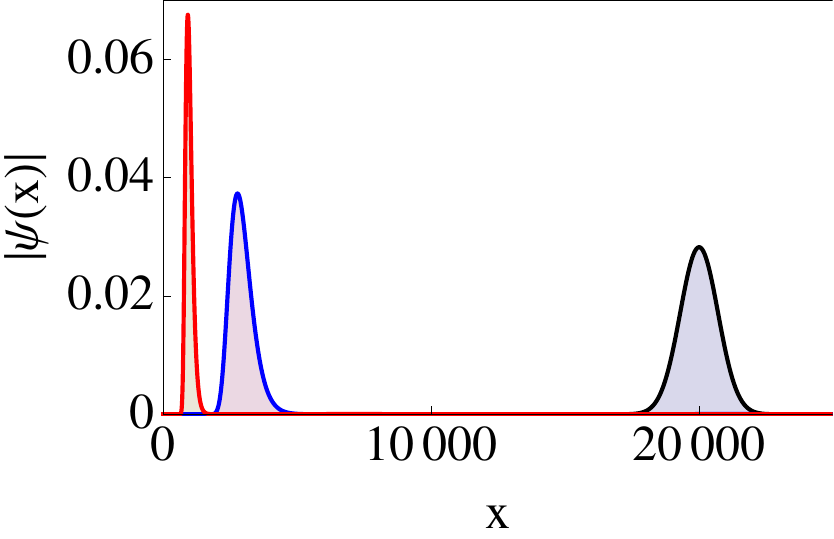} 
\caption{(Colors online) Transformation of a Gaussian state due to a boost for two
  different values of $\beta=-0.99,\,-0.999$ and $m=0.1$ in the
  momentum (left) and the position (right) representations.}
\label{fig:smooth}
\end{figure}

\begin{figure}[ht]
\includegraphics[width=\columnwidth]{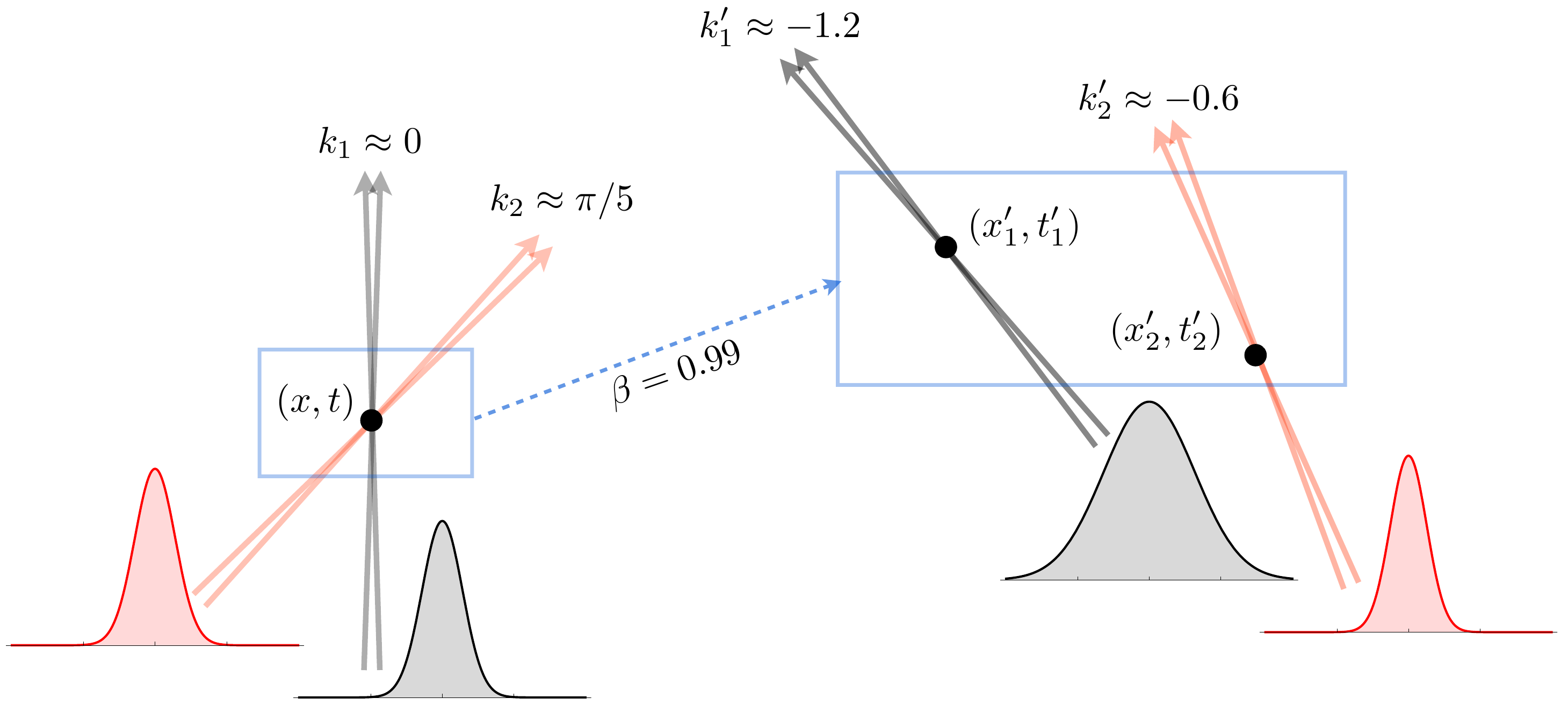}
\caption{(Colors online) Relative locality. In the left reference frame, the joint
  intersection of
four wave-packets, the first couple having wavevector close to $0$
and the second couple close to $\pi/5$, locates the point with
coordinates
$(x,t)$. In the boosted reference frame on the right, by applying the the transformation of
Eq. \eqref{eq:boost-automaton}, 
the  four
wave-packets no longer intersect at the same point.}\label{fig:collision2}
\end{figure}

Let us now deepen our analysis and consider how the features of the
present framework affect the geometry of space and time.  Under the
action of the deformed boost $\B$ a function $\hat{f}(\omega,k)$
transforms as $\hat{f}'(\omega,k) = \hat{f}(\omega'(\omega,k),k'(\omega,k))$ and,
following an ansatz due to Sch\"utzhold et al.
\cite{schutzhold2003large}, one can express the boosted function in
the variables $t,x$ by conjugating the boost $\B$ with the Fourier
transform $\mathcal{F}$ \footnote{Here $\mathcal{F}$ denotes the
  Fourier transform in both time and position variables: $\mathcal
  F(f)(\omega,k)=\sum_{t,x}e^{i(\omega t-kx)}f(\omega,k)$, $\mathcal
  F^{-1}(f)(t,x)=\int\!\mathrm{d}\mu \, e^{-i(\omega t-kx)}f(t,x)$}
\ie
\begin{align}
   & f'={\mathcal F}^{-1}\circ\B\circ {\mathcal F}\,f,\label{eq:ansatz}\\
   & f'(t',x')=\!\!\!\!\sum_{x,t\in{\mathbb Z}} \int\!\d\omega'\!\d k'\,
   e^{-i\chi(\omega',k',x,t,x',t')} f(x,t),\nonumber \\
   & \chi(\omega',k',x,t,x',t')=k(\omega',k')x-k' x' -\omega(\omega',k') t +\omega' t'.\nonumber
\end{align}
We notice that, due to the non linearity of $\M$, the map
\eqref{eq:ansatz} does not correspond to a change of coordinates from
$(t,x)$ to $(t',x')$ and therefore we cannot straightforwardly
interpret the variables $t$ and $x$ as the coordinates of points in a
continuum space-time interpolating the automaton cells: this may be regarded as manifestation of the
quantum nature of space-time.
\begin{figure}[ht]
  \includegraphics[width=\columnwidth]{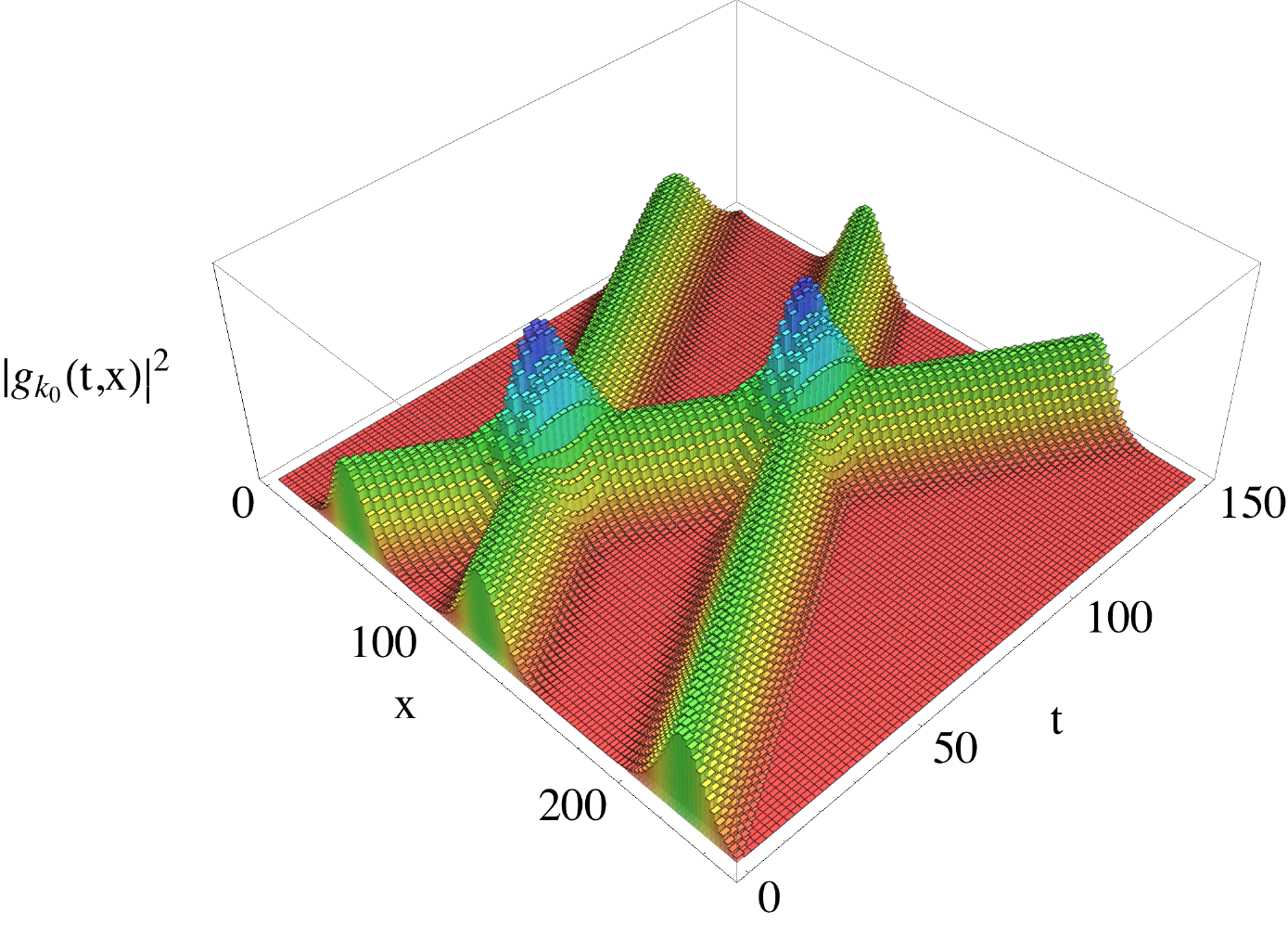}
  \caption{(Colors online) Two coincidences of travelling wave-packets in the automaton evolution of
    Eq. (\ref{eq:evol}) .}\label{fig:collision}
\end{figure}
One can then adopt the heuristic construction of
Ref.~\cite{schutzhold2003large}, interpreting physically the
coordinates $(x,t)$ in terms of the mean position $x$ at time $t$ of a
restricted class of states that can be interpreted as moving
particles, namely narrow-band Gaussian wave-packets moving at the
group-velocity. In this construction points in space-time are regarded
as crossing points of the trajectories of two particles (such points
have an ``extension'' due to the Gaussian profile). For a function
$g_{k_0}(t,x)$ peaked around $k_0$, the map \eqref{eq:ansatz} can be
approximated by taking the first order Taylor expansion of
$k(\omega',k')$ and $\omega(\omega',k')$ respectively around $k'_0$
and $\omega'(k_0)$ in the function $\chi$ (one can verify that a
narrow wave-packet remains narrow under a boost, see
Fig.~\ref{fig:smooth}, thus confirming the validity of the
approximation). This leads to the following transformations
  \begin{align}\label{eq:xttrans}
\begin{pmatrix}
t'\\x'
\end{pmatrix}
\approx
\begin{pmatrix}
-\partial_{\omega'} k&\partial_{k'} k\\
\partial_{\omega'} \omega&-\partial_{k'} \omega
\end{pmatrix}_{k'=k'_0}
\begin{pmatrix}
t\\x
\end{pmatrix}
\end{align}
Since Eq.~\eqref{eq:xttrans} defines a linear transformation of the variables $x$ and $t$ and the
wave-packets move along straight lines, we can interpret \eqref{eq:xttrans} as the transformation of
the coordinates $x_p$ and $t_p$ of a point $p$ in space-time, namely of the intersection of the
trajectories of two particles having $k$'s close to some common $k_0$.  However, the $k$-dependance
of the transformations \eqref{eq:xttrans} makes the geometry of space-time observer-dependent in the
following sense. Consider a point $p$ which is given by the intersection of four wave-packets, the
first pair peaked around $k_1$ and the second pair peaked around $k_2$ ($k_1\neq k_2$). Because of
the $k$ dependence in \eqref{eq:xttrans}, a boosted observer will actually see the first pair
intersecting at a point which is different from the one where the second pair intersects (see
Fig.~\ref{fig:collision2}).
This effect, first noticed in Ref.~\cite{schutzhold2003large} is the characteristic trait of the
so-called \emph{relative locality} \cite{amelino2011relative,PhysRevD.84.084010}. The space-time
resulting in such a way from the automaton dynamics is not ``objective'', in the sense that events
that coincide for one observer may not for another boosted observer. The above heuristic
construction is in agreement with the assertion of Ref. \cite{PhysRevD.84.084010} that relative
locality appears as a feature of all models in which the energy-momentum space has a non flat
geometry.  This can be easily seen by requiring that the transformation \eqref{eq:xttrans} does not
depend on $k_0$ and remembering that for $k_0=0$ one must recover the usual Lorentz transformations.
 
In this letter we have shown that the quantum cellular automaton of
Refs.~\cite{BDTqcaI,d2013derivation} provides a microscopic kinematical model compatible with the
recent proposals of DSR. We obtained the nonlinear representation of the Lorentz group in the
energy-momentum space by assuming the invariance of the dispersion relation of the automaton. Using
the aurguments of Ref. \cite{schutzhold2003large} we heuristically derived a space-time that
exhibits the phenomenon of relative locality. Our analysis has been carried in the easiest case of
one space dimension, which, however, is sufficient to the analysis of the present letter. The same
arguments can be easily generalized to three space dimensions using the results of Ref.
\cite{d2013derivation}, leading to additional symmetry violations, e.g. rotational covariance.


\acknowledgments This work has been supported in part by the Templeton
Foundation under the project ID\# 43796 {\em A Quantum-Digital
  Universe}. The work of Alexandre Bibeau-Delisle was supported in
part by the Natural Sciences and Engineering Research Council of
Canada through the Gerhard Herzberg Canada Gold Medal for Science and
Engineering.

\bibliography{bibliography}

\begin{thebibliography}{32}%
\makeatletter
\providecommand \@ifxundefined [1]{%
 \@ifx{#1\undefined}
}%
\providecommand \@ifnum [1]{%
 \ifnum #1\expandafter \@firstoftwo
 \else \expandafter \@secondoftwo
 \fi
}%
\providecommand \@ifx [1]{%
 \ifx #1\expandafter \@firstoftwo
 \else \expandafter \@secondoftwo
 \fi
}%
\providecommand \natexlab [1]{#1}%
\providecommand \enquote  [1]{``#1''}%
\providecommand \bibnamefont  [1]{#1}%
\providecommand \bibfnamefont [1]{#1}%
\providecommand \citenamefont [1]{#1}%
\providecommand \href@noop [0]{\@secondoftwo}%
\providecommand \href [0]{\begingroup \@sanitize@url \@href}%
\providecommand \@href[1]{\@@startlink{#1}\@@href}%
\providecommand \@@href[1]{\endgroup#1\@@endlink}%
\providecommand \@sanitize@url [0]{\catcode `\\12\catcode `\$12\catcode
  `\&12\catcode `\#12\catcode `\^12\catcode `\_12\catcode `\%12\relax}%
\providecommand \@@startlink[1]{}%
\providecommand \@@endlink[0]{}%
\providecommand \url  [0]{\begingroup\@sanitize@url \@url }%
\providecommand \@url [1]{\endgroup\@href {#1}{\urlprefix }}%
\providecommand \urlprefix  [0]{URL }%
\providecommand \Eprint [0]{\href }%
\providecommand \doibase [0]{http://dx.doi.org/}%
\providecommand \selectlanguage [0]{\@gobble}%
\providecommand \bibinfo  [0]{\@secondoftwo}%
\providecommand \bibfield  [0]{\@secondoftwo}%
\providecommand \translation [1]{[#1]}%
\providecommand \BibitemOpen [0]{}%
\providecommand \bibitemStop [0]{}%
\providecommand \bibitemNoStop [0]{.\EOS\space}%
\providecommand \EOS [0]{\spacefactor3000\relax}%
\providecommand \BibitemShut  [1]{\csname bibitem#1\endcsname}%
\let\auto@bib@innerbib\@empty
\bibitem [{\citenamefont {Polchinski}(1996)}]{polchinski1996tasi}%
  \BibitemOpen
  \bibfield  {author} {\bibinfo {author} {\bibfnamefont {J.}~\bibnamefont
  {Polchinski}},\ }\href@noop {} {\bibfield  {journal} {\bibinfo  {journal}
  {arXiv preprint hep-th/9611050}\ } (\bibinfo {year} {1996})}\BibitemShut
  {NoStop}%
\bibitem [{\citenamefont {Polchinski}(2003)}]{polchinski2003string}%
  \BibitemOpen
  \bibfield  {author} {\bibinfo {author} {\bibfnamefont {J.~G.}\ \bibnamefont
  {Polchinski}},\ }\href@noop {} {\emph {\bibinfo {title} {String theory}}}\
  (\bibinfo  {publisher} {Cambridge university press},\ \bibinfo {year}
  {2003})\BibitemShut {NoStop}%
\bibitem [{\citenamefont {Rovelli}\ and\ \citenamefont
  {Smolin}(1990)}]{rovelli1990loop}%
  \BibitemOpen
  \bibfield  {author} {\bibinfo {author} {\bibfnamefont {C.}~\bibnamefont
  {Rovelli}}\ and\ \bibinfo {author} {\bibfnamefont {L.}~\bibnamefont
  {Smolin}},\ }\href@noop {} {\bibfield  {journal} {\bibinfo  {journal}
  {Nuclear Physics B}\ }\textbf {\bibinfo {volume} {8}},\ \bibinfo {pages} {80}
  (\bibinfo {year} {1990})}\BibitemShut {NoStop}%
\bibitem [{\citenamefont {Ashtekar}\ \emph {et~al.}(1992)\citenamefont
  {Ashtekar}, \citenamefont {Rovelli},\ and\ \citenamefont
  {Smolin}}]{ashtekar1992weaving}%
  \BibitemOpen
  \bibfield  {author} {\bibinfo {author} {\bibfnamefont {A.}~\bibnamefont
  {Ashtekar}}, \bibinfo {author} {\bibfnamefont {C.}~\bibnamefont {Rovelli}}, \
  and\ \bibinfo {author} {\bibfnamefont {L.}~\bibnamefont {Smolin}},\
  }\href@noop {} {\bibfield  {journal} {\bibinfo  {journal} {Physical Review
  Letters}\ }\textbf {\bibinfo {volume} {69}},\ \bibinfo {pages} {237}
  (\bibinfo {year} {1992})}\BibitemShut {NoStop}%
\bibitem [{\citenamefont {Rovelli}\ and\ \citenamefont
  {Smolin}(1995)}]{rovelli1995discreteness}%
  \BibitemOpen
  \bibfield  {author} {\bibinfo {author} {\bibfnamefont {C.}~\bibnamefont
  {Rovelli}}\ and\ \bibinfo {author} {\bibfnamefont {L.}~\bibnamefont
  {Smolin}},\ }\href@noop {} {\bibfield  {journal} {\bibinfo  {journal}
  {Nuclear Physics B}\ }\textbf {\bibinfo {volume} {442}},\ \bibinfo {pages}
  {593} (\bibinfo {year} {1995})}\BibitemShut {NoStop}%
\bibitem [{\citenamefont {Carlip}(2001)}]{carlip2001quantum}%
  \BibitemOpen
  \bibfield  {author} {\bibinfo {author} {\bibfnamefont {S.}~\bibnamefont
  {Carlip}},\ }\href@noop {} {\bibfield  {journal} {\bibinfo  {journal}
  {Reports on progress in physics}\ }\textbf {\bibinfo {volume} {64}},\
  \bibinfo {pages} {885} (\bibinfo {year} {2001})}\BibitemShut {NoStop}%
\bibitem [{\citenamefont {Garay}(1995)}]{garay1995quantum}%
  \BibitemOpen
  \bibfield  {author} {\bibinfo {author} {\bibfnamefont {L.~J.}\ \bibnamefont
  {Garay}},\ }\href@noop {} {\bibfield  {journal} {\bibinfo  {journal}
  {International Journal of Modern Physics A}\ }\textbf {\bibinfo {volume}
  {10}},\ \bibinfo {pages} {145} (\bibinfo {year} {1995})}\BibitemShut
  {NoStop}%
\bibitem [{\citenamefont
  {Amelino-Camelia}(2002{\natexlab{a}})}]{amelino2002relativity}%
  \BibitemOpen
  \bibfield  {author} {\bibinfo {author} {\bibfnamefont {G.}~\bibnamefont
  {Amelino-Camelia}},\ }\href@noop {} {\bibfield  {journal} {\bibinfo
  {journal} {International Journal of Modern Physics D}\ }\textbf {\bibinfo
  {volume} {11}},\ \bibinfo {pages} {35} (\bibinfo {year}
  {2002}{\natexlab{a}})}\BibitemShut {NoStop}%
\bibitem [{\citenamefont {Amelino-Camelia}\ and\ \citenamefont
  {Piran}(2001)}]{amelino2001planck}%
  \BibitemOpen
  \bibfield  {author} {\bibinfo {author} {\bibfnamefont {G.}~\bibnamefont
  {Amelino-Camelia}}\ and\ \bibinfo {author} {\bibfnamefont {T.}~\bibnamefont
  {Piran}},\ }\href@noop {} {\bibfield  {journal} {\bibinfo  {journal}
  {Physical Review D}\ }\textbf {\bibinfo {volume} {64}},\ \bibinfo {pages}
  {036005} (\bibinfo {year} {2001})}\BibitemShut {NoStop}%
\bibitem [{\citenamefont
  {Amelino-Camelia}(2002{\natexlab{b}})}]{amelino2002quantum}%
  \BibitemOpen
  \bibfield  {author} {\bibinfo {author} {\bibfnamefont {G.}~\bibnamefont
  {Amelino-Camelia}},\ }\href@noop {} {\bibfield  {journal} {\bibinfo
  {journal} {Modern Physics Letters A}\ }\textbf {\bibinfo {volume} {17}},\
  \bibinfo {pages} {899} (\bibinfo {year} {2002}{\natexlab{b}})}\BibitemShut
  {NoStop}%
\bibitem [{\citenamefont {Amelino-Camelia}(2001)}]{amelino2001testable}%
  \BibitemOpen
  \bibfield  {author} {\bibinfo {author} {\bibfnamefont {G.}~\bibnamefont
  {Amelino-Camelia}},\ }\href@noop {} {\bibfield  {journal} {\bibinfo
  {journal} {Physics Letters B}\ }\textbf {\bibinfo {volume} {510}},\ \bibinfo
  {pages} {255} (\bibinfo {year} {2001})}\BibitemShut {NoStop}%
\bibitem [{\citenamefont {Magueijo}\ and\ \citenamefont
  {Smolin}(2002)}]{PhysRevLett.88.190403}%
  \BibitemOpen
  \bibfield  {author} {\bibinfo {author} {\bibfnamefont {J.~a.}\ \bibnamefont
  {Magueijo}}\ and\ \bibinfo {author} {\bibfnamefont {L.}~\bibnamefont
  {Smolin}},\ }\href {\doibase 10.1103/PhysRevLett.88.190403} {\bibfield
  {journal} {\bibinfo  {journal} {Phys. Rev. Lett.}\ }\textbf {\bibinfo
  {volume} {88}},\ \bibinfo {pages} {190403} (\bibinfo {year}
  {2002})}\BibitemShut {NoStop}%
\bibitem [{\citenamefont {Magueijo}\ and\ \citenamefont
  {Smolin}(2003)}]{magueijo2003generalized}%
  \BibitemOpen
  \bibfield  {author} {\bibinfo {author} {\bibfnamefont {J.}~\bibnamefont
  {Magueijo}}\ and\ \bibinfo {author} {\bibfnamefont {L.}~\bibnamefont
  {Smolin}},\ }\href@noop {} {\bibfield  {journal} {\bibinfo  {journal}
  {Physical Review D}\ }\textbf {\bibinfo {volume} {67}},\ \bibinfo {pages}
  {044017} (\bibinfo {year} {2003})}\BibitemShut {NoStop}%
\bibitem [{\citenamefont {Takeda}\ \emph {et~al.}(1998)\citenamefont {Takeda},
  \citenamefont {Hayashida}, \citenamefont {Honda}, \citenamefont {Inoue},
  \citenamefont {Kadota}, \citenamefont {Kakimoto}, \citenamefont {Kamata},
  \citenamefont {Kawaguchi}, \citenamefont {Kawasaki}, \citenamefont {Kawasumi}
  \emph {et~al.}}]{takeda1998extension}%
  \BibitemOpen
  \bibfield  {author} {\bibinfo {author} {\bibfnamefont {M.}~\bibnamefont
  {Takeda}}, \bibinfo {author} {\bibfnamefont {N.}~\bibnamefont {Hayashida}},
  \bibinfo {author} {\bibfnamefont {K.}~\bibnamefont {Honda}}, \bibinfo
  {author} {\bibfnamefont {N.}~\bibnamefont {Inoue}}, \bibinfo {author}
  {\bibfnamefont {K.}~\bibnamefont {Kadota}}, \bibinfo {author} {\bibfnamefont
  {F.}~\bibnamefont {Kakimoto}}, \bibinfo {author} {\bibfnamefont
  {K.}~\bibnamefont {Kamata}}, \bibinfo {author} {\bibfnamefont
  {S.}~\bibnamefont {Kawaguchi}}, \bibinfo {author} {\bibfnamefont
  {Y.}~\bibnamefont {Kawasaki}}, \bibinfo {author} {\bibfnamefont
  {N.}~\bibnamefont {Kawasumi}},  \emph {et~al.},\ }\href@noop {} {\bibfield
  {journal} {\bibinfo  {journal} {Physical Review Letters}\ }\textbf {\bibinfo
  {volume} {81}},\ \bibinfo {pages} {1163} (\bibinfo {year}
  {1998})}\BibitemShut {NoStop}%
\bibitem [{\citenamefont {Finkbeiner}\ \emph {et~al.}(2000)\citenamefont
  {Finkbeiner}, \citenamefont {Davis},\ and\ \citenamefont
  {Schlegel}}]{finkbeiner2000detection}%
  \BibitemOpen
  \bibfield  {author} {\bibinfo {author} {\bibfnamefont {D.~P.}\ \bibnamefont
  {Finkbeiner}}, \bibinfo {author} {\bibfnamefont {M.}~\bibnamefont {Davis}}, \
  and\ \bibinfo {author} {\bibfnamefont {D.~J.}\ \bibnamefont {Schlegel}},\
  }\href@noop {} {\bibfield  {journal} {\bibinfo  {journal} {The Astrophysical
  Journal}\ }\textbf {\bibinfo {volume} {544}},\ \bibinfo {pages} {81}
  (\bibinfo {year} {2000})}\BibitemShut {NoStop}%
\bibitem [{\citenamefont {Amelino-Camelia}\ \emph {et~al.}(2009)\citenamefont
  {Amelino-Camelia}, \citenamefont {L\"ammerzahl}, \citenamefont {Mercati},\
  and\ \citenamefont {Tino}}]{PhysRevLett.103.171302}%
  \BibitemOpen
  \bibfield  {author} {\bibinfo {author} {\bibfnamefont {G.}~\bibnamefont
  {Amelino-Camelia}}, \bibinfo {author} {\bibfnamefont {C.}~\bibnamefont
  {L\"ammerzahl}}, \bibinfo {author} {\bibfnamefont {F.}~\bibnamefont
  {Mercati}}, \ and\ \bibinfo {author} {\bibfnamefont {G.~M.}\ \bibnamefont
  {Tino}},\ }\href {\doibase 10.1103/PhysRevLett.103.171302} {\bibfield
  {journal} {\bibinfo  {journal} {Phys. Rev. Lett.}\ }\textbf {\bibinfo
  {volume} {103}},\ \bibinfo {pages} {171302} (\bibinfo {year}
  {2009})}\BibitemShut {NoStop}%
\bibitem [{\citenamefont {D'Ariano}(2010)}]{veryoriginal}%
  \BibitemOpen
  \bibfield  {author} {\bibinfo {author} {\bibfnamefont {G.~M.}\ \bibnamefont
  {D'Ariano}},\ }in\ \href@noop {} {\emph {\bibinfo {booktitle} {Quantum
  Theory: Reconsideration of Foundations, 5}}},\ Vol.\ \bibinfo {volume}
  {CP1232}\ (\bibinfo {organization} {AIP},\ \bibinfo {year} {2010})\
  p.~\bibinfo {pages} {3},\ \bibinfo {note} {(arXiv:1001.1088)}\BibitemShut
  {NoStop}%
\bibitem [{\citenamefont {D'Ariano}(2011)}]{darianopla}%
  \BibitemOpen
  \bibfield  {author} {\bibinfo {author} {\bibfnamefont {G.~M.}\ \bibnamefont
  {D'Ariano}},\ }\href@noop {} {\bibfield  {journal} {\bibinfo  {journal}
  {Phys. Lett. A}\ }\textbf {\bibinfo {volume} {376}} (\bibinfo {year}
  {2011})}\BibitemShut {NoStop}%
\bibitem [{\citenamefont {Bisio}\ \emph {et~al.}(2012)\citenamefont {Bisio},
  \citenamefont {D'Ariano},\ and\ \citenamefont {Tosini}}]{BDTqcaI}%
  \BibitemOpen
  \bibfield  {author} {\bibinfo {author} {\bibfnamefont {A.}~\bibnamefont
  {Bisio}}, \bibinfo {author} {\bibfnamefont {G.}~\bibnamefont {D'Ariano}}, \
  and\ \bibinfo {author} {\bibfnamefont {A.}~\bibnamefont {Tosini}},\
  }\href@noop {} {\bibfield  {journal} {\bibinfo  {journal} {arXiv preprint
  arXiv:1212.2839}\ } (\bibinfo {year} {2012})}\BibitemShut {NoStop}%
\bibitem [{\citenamefont {D'Ariano}\ and\ \citenamefont
  {Perinotti}(2013)}]{d2013derivation}%
  \BibitemOpen
  \bibfield  {author} {\bibinfo {author} {\bibfnamefont {G.~M.}\ \bibnamefont
  {D'Ariano}}\ and\ \bibinfo {author} {\bibfnamefont {P.}~\bibnamefont
  {Perinotti}},\ }\href@noop {} {\bibfield  {journal} {\bibinfo  {journal}
  {arXiv preprint arXiv:1306.1934}\ } (\bibinfo {year} {2013})}\BibitemShut
  {NoStop}%
\bibitem [{\citenamefont {Bisio}\ \emph {et~al.}(2013)\citenamefont {Bisio},
  \citenamefont {D'Ariano},\ and\ \citenamefont {Tosini}}]{bisio2013dirac}%
  \BibitemOpen
  \bibfield  {author} {\bibinfo {author} {\bibfnamefont {A.}~\bibnamefont
  {Bisio}}, \bibinfo {author} {\bibfnamefont {G.~M.}\ \bibnamefont {D'Ariano}},
  \ and\ \bibinfo {author} {\bibfnamefont {A.}~\bibnamefont {Tosini}},\ }\href
  {\doibase 10.1103/PhysRevA.88.032301} {\bibfield  {journal} {\bibinfo
  {journal} {Phys. Rev. A}\ }\textbf {\bibinfo {volume} {88}},\ \bibinfo
  {pages} {032301} (\bibinfo {year} {2013})}\BibitemShut {NoStop}%
\bibitem [{\citenamefont {von Neumann}(1966)}]{neumann1966theory}%
  \BibitemOpen
  \bibfield  {author} {\bibinfo {author} {\bibfnamefont {J.}~\bibnamefont {von
  Neumann}},\ }\href@noop {} {\emph {\bibinfo {title} {Theory of
  self-reproducing automata}}}\ (\bibinfo  {publisher} {University of Illinois
  Press},\ \bibinfo {address} {Urbana and London},\ \bibinfo {year}
  {1966})\BibitemShut {NoStop}%
\bibitem [{Note1()}]{Note1}%
  \BibitemOpen
  \bibinfo {note} {The general theory of QCA is rigorously treated in Refs.
  \cite {arrighi2011unitarity,gross2012index} while first attempts to model
  relativistic dynamics with QCA appears in \cite
  {bialynicki1994weyl,meyer1996quantum}.}\BibitemShut {Stop}%
\bibitem [{\citenamefont {Sch{\"u}tzhold}\ and\ \citenamefont
  {Unruh}(2003)}]{schutzhold2003large}%
  \BibitemOpen
  \bibfield  {author} {\bibinfo {author} {\bibfnamefont {R.}~\bibnamefont
  {Sch{\"u}tzhold}}\ and\ \bibinfo {author} {\bibfnamefont {W.}~\bibnamefont
  {Unruh}},\ }\href@noop {} {\bibfield  {journal} {\bibinfo  {journal} {Journal
  of Experimental and Theoretical Physics Letters}\ }\textbf {\bibinfo {volume}
  {78}},\ \bibinfo {pages} {431} (\bibinfo {year} {2003})}\BibitemShut
  {NoStop}%
\bibitem [{\citenamefont {Amelino-Camelia}\ \emph
  {et~al.}(2011{\natexlab{a}})\citenamefont {Amelino-Camelia}, \citenamefont
  {Freidel}, \citenamefont {Kowalski-Glikman},\ and\ \citenamefont
  {Smolin}}]{amelino2011relative}%
  \BibitemOpen
  \bibfield  {author} {\bibinfo {author} {\bibfnamefont {G.}~\bibnamefont
  {Amelino-Camelia}}, \bibinfo {author} {\bibfnamefont {L.}~\bibnamefont
  {Freidel}}, \bibinfo {author} {\bibfnamefont {J.}~\bibnamefont
  {Kowalski-Glikman}}, \ and\ \bibinfo {author} {\bibfnamefont
  {L.}~\bibnamefont {Smolin}},\ }\href@noop {} {\bibfield  {journal} {\bibinfo
  {journal} {International Journal of Modern Physics D}\ }\textbf {\bibinfo
  {volume} {20}},\ \bibinfo {pages} {2867} (\bibinfo {year}
  {2011}{\natexlab{a}})}\BibitemShut {NoStop}%
\bibitem [{\citenamefont {Amelino-Camelia}\ \emph
  {et~al.}(2011{\natexlab{b}})\citenamefont {Amelino-Camelia}, \citenamefont
  {Freidel}, \citenamefont {Kowalski-Glikman},\ and\ \citenamefont
  {Smolin}}]{PhysRevD.84.084010}%
  \BibitemOpen
  \bibfield  {author} {\bibinfo {author} {\bibfnamefont {G.}~\bibnamefont
  {Amelino-Camelia}}, \bibinfo {author} {\bibfnamefont {L.}~\bibnamefont
  {Freidel}}, \bibinfo {author} {\bibfnamefont {J.}~\bibnamefont
  {Kowalski-Glikman}}, \ and\ \bibinfo {author} {\bibfnamefont
  {L.}~\bibnamefont {Smolin}},\ }\href {\doibase 10.1103/PhysRevD.84.084010}
  {\bibfield  {journal} {\bibinfo  {journal} {Phys. Rev. D}\ }\textbf {\bibinfo
  {volume} {84}},\ \bibinfo {pages} {084010} (\bibinfo {year}
  {2011}{\natexlab{b}})}\BibitemShut {NoStop}%
\bibitem [{\citenamefont {Meyer}(1996)}]{meyer1996quantum}%
  \BibitemOpen
  \bibfield  {author} {\bibinfo {author} {\bibfnamefont {D.}~\bibnamefont
  {Meyer}},\ }\href@noop {} {\bibfield  {journal} {\bibinfo  {journal} {Journal
  of Statistical Physics}\ }\textbf {\bibinfo {volume} {85}},\ \bibinfo {pages}
  {551} (\bibinfo {year} {1996})}\BibitemShut {NoStop}%
\bibitem [{Note2()}]{Note2}%
  \BibitemOpen
  \bibinfo {note} {For large wave-vectors a thorough interpretation of $k$ and
  $\omega $ in terms of the momentum and energy, respectively, would need a
  full interacting theory.}\BibitemShut {Stop}%
\bibitem [{Note3()}]{Note3}%
  \BibitemOpen
  \bibinfo {note} {Here $\protect \mathcal {F}$ denotes the Fourier transform
  in both time and position variables: $\protect \mathcal F(f)(\omega
  ,k)=\DOTSB \sum@ \slimits@ _{t,x}e^{i(\omega t-kx)}f(\omega ,k)$, $\protect
  \mathcal F^{-1}(f)(t,x)=\DOTSI \intop \ilimits@ \protect \tmspace
  -\thinmuskip {.1667em}\protect \mathrm {d}\mu \protect \tmspace +\thinmuskip
  {.1667em} e^{-i(\omega t-kx)}f(t,x)$}\BibitemShut {NoStop}%
\bibitem [{\citenamefont {Arrighi}\ \emph {et~al.}(2011)\citenamefont
  {Arrighi}, \citenamefont {Nesme},\ and\ \citenamefont
  {Werner}}]{arrighi2011unitarity}%
  \BibitemOpen
  \bibfield  {author} {\bibinfo {author} {\bibfnamefont {P.}~\bibnamefont
  {Arrighi}}, \bibinfo {author} {\bibfnamefont {V.}~\bibnamefont {Nesme}}, \
  and\ \bibinfo {author} {\bibfnamefont {R.}~\bibnamefont {Werner}},\
  }\href@noop {} {\bibfield  {journal} {\bibinfo  {journal} {Journal of
  Computer and System Sciences}\ }\textbf {\bibinfo {volume} {77}},\ \bibinfo
  {pages} {372} (\bibinfo {year} {2011})}\BibitemShut {NoStop}%
\bibitem [{\citenamefont {Gross}\ \emph {et~al.}(2012)\citenamefont {Gross},
  \citenamefont {Nesme}, \citenamefont {Vogts},\ and\ \citenamefont
  {Werner}}]{gross2012index}%
  \BibitemOpen
  \bibfield  {author} {\bibinfo {author} {\bibfnamefont {D.}~\bibnamefont
  {Gross}}, \bibinfo {author} {\bibfnamefont {V.}~\bibnamefont {Nesme}},
  \bibinfo {author} {\bibfnamefont {H.}~\bibnamefont {Vogts}}, \ and\ \bibinfo
  {author} {\bibfnamefont {R.}~\bibnamefont {Werner}},\ }\href@noop {}
  {\bibfield  {journal} {\bibinfo  {journal} {Communications in Mathematical
  Physics}\ ,\ \bibinfo {pages} {1}} (\bibinfo {year} {2012})}\BibitemShut
  {NoStop}%
\bibitem [{\citenamefont {Bialynicki-Birula}(1994)}]{bialynicki1994weyl}%
  \BibitemOpen
  \bibfield  {author} {\bibinfo {author} {\bibfnamefont {I.}~\bibnamefont
  {Bialynicki-Birula}},\ }\href@noop {} {\bibfield  {journal} {\bibinfo
  {journal} {Physical Review D}\ }\textbf {\bibinfo {volume} {49}},\ \bibinfo
  {pages} {6920} (\bibinfo {year} {1994})}\BibitemShut {NoStop}%
\end{thebibliography}%

\end{document}